\documentclass[twocolumn,showpacs,preprintnumbers,amsmath,amssymb]{revtex4}

\usepackage{graphicx}
\usepackage{dcolumn}
\usepackage{bm}

\begin{document}

\preprint{APS/123-QED}

\title{Effect of Elastic Deformations on the Multicritical Behavior
of Disordered Systems}

\author{S.V. Belim}
 \email{belim@univer.omsk.su}
\affiliation{%
Omsk State University, 55-a, pr. Mira, Omsk, Russia, 644077
\textbackslash\textbackslash
}%

\date{\today}

\begin{abstract}
A field-theoretical description of the behavior of disordered,
elastically isotropic, compressible systems characterized by two
order parameters at the bicritical and tetracritical points is
presented. The description is performed in the two-loop
approximation in three dimensions . The renormalization group
equations are analyzed, and the fixed points corresponding to
different types of multicritical behavior are determined. It is
shown that the effect of elastic deformations causes a change in
the regime of the tetracritical behavior of disordered systems
because of the interaction of the order parameters through the
deformation field.
\end{abstract}

\pacs{64.60.-i}
\maketitle

It has been shown in [1] that the presence of frozen structural
point defects gives rise to changes in the behavior of a system in
both bicritical and tetracritical regions. According to the cited
paper, the effect of $\delta$-correlated impurities is essential
only for Ising systems and leads to a decoupling of the order
parameters at the multicritical points. Later [2], it was shown
that elastic deformations cause a renormalization of the effective
charges for the interaction of critical fluctuations, which lead
to an increase in the interaction of order parameters and to a
change in the type of multicritical behavior. Therefore, the study
of the combined effect of elastic deformations and structural
point defects in the multicritical region is of great interest.

The purpose of this paper is to study the influence of striction
effects on disordered systems whose phase diagrams already contain
multicritical points of bicritical or tetracritical type. In the
first case, a multicritical point is the point of intersection of
two lines of second-order phase transitions and one line of
first-order phase transitions, while in the second case, it
corresponds to the intersection of four lines of second-order
phase transitions. In the immediate vicinity of a multicritical
point, the system exhibits a specific critical behavior
characterized by the competition between the types of ordering. At
a bicritical point, one parameter is displaced by another, whereas
the tetracritical point allows a mixed phase with a coexistence of
different types of ordering. Such systems [1] can be described by
introducing two order parameters that belong to different
irreducible representations.

In structural phase transitions that occur in the absence of
piezoelectric effect, in the paraphase, elastic strains play the
role of a secondary order parameter whose fluctuations are not
critical in most cases [3, 4]. Since, in the critical region, the
main contribution to the striction effects comes from the
dependence of the exchange integral on distance, only elastically
isotropic systems are considered below.

The model Hamiltonian of a system has the form
\begin{eqnarray}\label{gam1}
&&H_0=\int d^Dx\Big[\frac{1}{2}(\tau_1+\nabla^2)\sum_{a=1}^{m}\Phi^a(x)^{2}\\
&&+\frac{1}{2}(\tau_1+\nabla^2)\sum_{a=1}^{m}\Psi^a(x)^{2}
+\frac{u_{01}}{4!}\sum_{a,b=1}^{m}\Phi^a(x)^{2}\Phi^b(x)^{2}\nonumber\\
&&+\frac{u_{02}}{4!}\sum_{a,b=1}^{m}\Psi^a(x)^{2}\Psi^b(x)^{2}
+\frac{2u_{03}}{4!}\sum_{a,b=1}^{m}(\Phi^a(x)\Psi^b(x))^2\nonumber\\
&&-\frac{\delta_{01}}{2}\sum_{a=1}^{m}\Phi^a(x)^{4}
-\frac{\delta_{02}}{2}\sum_{a=1}^{m}\Psi^a(x)^{4}\nonumber\\
&&-\delta_{03}\sum_{a=1}^{m}\Phi^a(x)^{2}\Psi^a(x)^{2}
+g_1y(x)\sum_{a=1}^{m}\Phi^a(x)^{2}\nonumber\\
&&+g_2y(x)\sum_{a=1}^{m}\Psi^a(x)^{2}
+\beta y(x)^2\Big],\nonumber
\end{eqnarray}
Here, $\Phi(x)$ and $\Psi(x)$ are the fluctuating order parameters;
$u_{01}$ and $u_{02}$ are positive constants; $\tau_1\sim|T-T_{c1}|/T_{c1}$
and $\tau_2\sim|T-T_{c2}|/T_{c2}$, where $T_{c1}$ and $T_{c2}$ are the
phase transition temperatures for the first- and second-order
parameters, respectively; $y(x)=\sum\limits_{\alpha=1}^3u_{\alpha\alpha}(x)$
where $u_{\alpha \beta}$ is the strain
tensor; $g_1$ and $g_2$ are the quadratic striction parameters;
$\beta$ is a constant characterizing the elastic properties of the crystal;
and $D$ is the space dimension. In this Hamiltonian, the integration
with respect to the components depending on the nonfluctuating
variables, which do not interact with the order parameters, is
already performed along with the replica procedure of averaging
over impurities. The properties of the initial system can be
obtained in the limit $m \rightarrow 0$. The nonnegative constants
$\delta_{01}$, $\delta_{02}$ and $\delta_{03}$
describe the interaction of critical fluctuations via the
impurity field. The interaction of impurities with elastic
deformations has linear character and, in the case of averaging
over impurities, leads to the redetermination of the constants
$\delta_{01}$, $\delta_{02}$ and $\delta_{03}$ [5].

By going to the Fourier transforms of the variables in Eq. (1), we
obtain the Hamiltonian of the system in the form
\begin{eqnarray}\label{gam2}
&&H_0=\frac 12\int d^Dq(\tau _1+q^2)\sum_{a=1}^{m}\Phi_q^{a}\Phi_{-q}^{a}\nonumber\\
&&+\frac 12\int d^Dq(\tau _2+q^2)\sum_{a=1}^{m}\Psi_q^{a}\Psi_{-q}^{a}\\
&&+\frac{u_{01}}{4!}\int d^D{q_i}\sum_{a,b=1}^{m}(\Phi_{q1}^{a}\Phi_{q2}^{a})(\Phi_{q3}^{b}\Phi_{-q1-q2-q3}^{b})\nonumber\\
&&+\frac{u_{02}}{4!}\int d^D{q_i}\sum_{a,b=1}^{m}(\Psi_{q1}^{a}\Psi_{q2}^{a})(\Psi_{q3}^{b}\Psi_{-q1-q2-q3}^{b})\nonumber\\
&&+\frac{2u_{03}}{4!}\int d^D{q_i}\sum_{a,b=1}^{m}(\Phi_{q1}^{a}\Phi_{q2}^{a})(\Psi_{q3}^{b}\Psi_{-q1-q2-q3}^{b})\nonumber\\
&&-\frac{\delta_{01}}{2}\int d^D{q_i}\sum_{a=1}^{m}(\Phi_{q1}^{a}\Phi_{q2}^{a})(\Phi_{q3}^{a}\Phi_{-q1-q2-q3}^{a})\nonumber\\
&&-\frac{\delta_{02}}{2}\int d^D{q_i}\sum_{a=1}^{m}(\Psi_{q1}^{a}\Psi_{q2}^{a})(\Psi_{q3}^{a}\Psi_{-q1-q2-q3}^{a})\nonumber\\
&&-\delta_{03}\int d^D{q_i}\sum_{a=1}^{m}(\Phi_{q1}^{a}\Phi_{q2}^{a})(\Psi_{q3}^{a}\Psi_{-q1-q2-q3}^{a})\nonumber\\
&&+g_1\int d^Dqy_{q1}\sum_{a=1}^{m}\Phi_{q2}^{a}\Phi_{-q1-q2}^{a}\nonumber\\
&&+g_2\int d^Dqy_{q1}\sum_{a=1}^{m}\Psi_{q2}^{a}\Psi_{-q1-q2}^{a}\nonumber\\
&&+\frac{g^0_1}{\Omega}y_0\int d^Dq\sum_{a=1}^{m}\Phi_{q}^{a}\Phi_{-q}^{a}\nonumber\\
&&+\frac{g^0_2}{\Omega}y_0\int d^Dq\sum_{a=1}^{m}\Psi_{q}^{a}\Psi_{-q}^{a}
+2\beta\int d^Dq y_qy_{-q} +2\frac{\beta_0}{\Omega}y_0^2\nonumber
\end{eqnarray}
Here, the components $y_0$ describing uniform strains are separated.
According to [3], such a separation is necessary, because the
nonuniform strains $y_q$ are responsible for the exchange of acoustic
phonons and lead to long-range interactions, which are absent in
the case of uniform deformations.

We define the effective Hamiltonian that depends only on the
strongly fluctuating order parameters $\Phi$ and $\Psi$of the system as
follows:
\begin{eqnarray}\label{3}
\exp \{-H[\Phi,\Psi]\}=B\int \exp \{-H_{0}[\Phi,\Psi,y]\}\prod dy_q
\end{eqnarray}
If the experiment is performed at a constant volume, the quantity
$y_0$ is a constant, and the integration in Eq. (3) is performed with
respect to only the nonuniform strains, while the uniform strains
do not contribute to the effective Hamiltonian. In an experiment
at a constant pressure, the term $P\Omega$ is added to the Hamiltonian,
with the volume being represented in terms of the strain tensor
components in the form
\begin{eqnarray}\label{6_1}
\Omega=\Omega_0 [1+\sum\limits_{\alpha =1}u_{\alpha\alpha}+
\sum\limits_{\alpha \neq \beta}u_{\alpha\alpha}u_{\beta\beta}+O(u^3)]
\end{eqnarray}
and the integration in Eq. (3) is performed also with respect to
the uniform strains. According to [6], the inclusion of quadratic
terms in Eq. (4) may be important at high pressures and for
crystals with strong striction effects. As a result, we obtain
\begin{eqnarray}\label{gam3}
&&H =\frac 12\int d^Dq(\tau _1+q^2)\sum_{a=1}^{m}\Phi_q^a\Phi_{-q}^a\\
&&+\frac 12\int d^Dq(\tau _2+q^2)\sum_{a=1}^{m}\Psi_q^a\Psi_{-q}^a\nonumber\\
&&+\frac{v_{01}}{4!}\int d^D{q_i}\sum_{a,b=1}^{m}(\Phi_{q1}^a\Phi_{q2}^a)(\Phi_{q3}^b\Phi_{-q1-q2-q3}^b)\nonumber\\
&&+\frac{v_{02}}{4!}\int d^D{q_i}\sum_{a,b=1}^{m}(\Psi_{q1}^a\Psi_{q2}^a)(\Psi_{q3}^b\Psi_{-q1-q2-q3}^b)\nonumber\\
&&+\frac{2v_{03}}{4!}\int d^D{q_i}\sum_{a,b=1}^{m}(\Phi_{q1}^a\Phi_{q2}^a)(\Psi_{q3}^b\Psi_{-q1-q2-q3}^b)\nonumber\\
&&-\frac{\delta_{01}}{2}\int d^D{q_i}\sum_{a=1}^{m}(\Phi_{q1}^a\Phi_{q2}^a)(\Phi_{q3}^a\Phi_{-q1-q2-q3}^a)\nonumber\\
&&-\frac{\delta_{02}}{2}\int d^D{q_i}\sum_{a=1}^{m}(\Psi_{q1}^a\Psi_{q2}^a)(\Psi_{q3}^a\Psi_{-q1-q2-q3}^a)\nonumber\\
&&-\delta_{03}\int d^D{q_i}\sum_{a=1}^{m}(\Phi_{q1}^a\Phi_{q2}^a)(\Psi_{q3}^a\Psi_{-q1-q2-q3}^a)\nonumber\\
&&+\frac{z_1^2-w_1^2}{2}\int d^D{q_i}\sum_{a,b=1}^{m}(\Phi_{q1}^a\Phi_{-q1}^a)(\Phi_{q2}^b\Phi_{-q2}^b)\nonumber\\
&&+\frac{z_2^2-w_2^2}{2}\int d^D{q_i}\sum_{a=1}^{m}(\Psi_{q1}^a\Psi_{-q1}^a)(\Psi_{q2}^a\Psi_{-q2}^a)\nonumber\\
&&+(z_1z_2-w_1w_2)\int d^D{q_i}\sum_{a,b=1}^{m}(\Phi_{q1}^a\Phi_{-q1}^a)(\Psi_{q2}^b\Psi_{-q2}^b)\nonumber \\
&& v_{01}=u_{01}-12z_1^2, \ \ v_{02}=u_{02}-12z_2^2, \nonumber\\
&&v_{03}=u_{03}-12z_1z_2,\nonumber \\
&& z_1=\frac{g_1}{\sqrt\beta},\ \ z_2=\frac{g_2}{\sqrt\beta},\ \
w_1=\frac{g^0_1}{\sqrt\beta_0},\ \ w_2=\frac{g^0_2}{\sqrt\beta_0}\nonumber
\end{eqnarray}
This Hamiltonian leads to a wide variety of multicritical points.
As for incompressible systems, both tetracritical
$(v_3+12(z_1z_2-w_1w_2))^2<(v_1+12(z_1^2-w_1^2))(v_2+12(z_2^2-w_2^2))$\\
and bicritical\\
$(v_3+12(z_1z_2-w_1w_2))^2\geq(v_1+12(z_1^2-w_1^2))(v_2+12(z_2^2-w_2^2))$\\
behaviors are possible. In addition, the striction effects may
give rise to multicritical points of higher orders.

In the framework of the field-theoretical approach [7], the
asymptotic critical behavior and the structure of the phase
diagram in the fluctuation region are determined by the
Callan–Symanzik renormalization-group equation for the vertex
parts of the irreducible Green's functions. To calculate the
$\beta$ and $\gamma$-functions as functions involved in the
Callan–Symanzik equation for renormalized interaction vertices
$u_1, u_2, u_3, \delta_1, \delta_2, \delta_2, g_1, g_2, g_1^{(0)}, g_2^{(0)}$
or for complex vertices
$z_1$, $z_2$,  $w_1$, $w_2$, $v_1$, $v_2$, $v_3$, $\delta_1$, $\delta_2$,
$\delta_3$, which are more convenient for the determination of the
multicritical behavior, the standard method based on the Feynman
diagram technique and on the renormalization procedure was used
[8]. As a result, the following expressions were obtained for the
â functions in the two-loop approximation:
\begin{eqnarray}\label{8}
&&\beta _{v1}=-v_1+\frac{3}{2}v_1^2+\frac{1}{6}v_3^2-24v_1\delta_1-\frac{77}{81}v_1^3\\
&&-\frac{23}{243}v_1v_3^2-\frac{2}{27}v_3^3
+\frac{184}{81}v_1v_3\delta_3+\frac{16}{9}v_3^2\delta_3\nonumber\\
&&+\frac{832}{27}v_1^2\delta_1-\frac{5920}{27}v_1\delta_1^2+\frac{8}{9}v_3^2\delta_1,\nonumber\\
&&\beta _{v2}=-v_2+\frac{3}{2}v_2^2+\frac{1}{6}v_3^2-24v_2\delta_1-\frac{77}{81}v_2^3\nonumber\\
&&-\frac{23}{243}v_2v_3^2-\frac{2}{27}v_3^3
+\frac{184}{81}v_2v_3\delta_3+\frac{16}{9}v_3^2\delta_3+\nonumber\\
&&+\frac{832}{27}v_2^2\delta_2
-\frac{5920}{27}v_2\delta_2^2+\frac{8}{9}v_3^2\delta_2,\nonumber\\
&&\beta_{v3}=-v_3+\frac{2}{3}v_3^2+\frac{1}{2}v_1v_3+\frac{1}{2}v_2v_3-4v_3\delta_1\nonumber\\
&&-4v_3\delta_2-16v_3\delta_3
-\frac{41}{243}v_3^3-\frac{23}{162}v_1^2v_3-\nonumber\\
&&-\frac{23}{162}v_2^2v_3-\frac{1}{3}v_1v_3^2-\frac{1}{3}v_2v_3^2
+\frac{472}{81}v_3^2\delta_3+\frac{8}{3}v_3^2\delta_1\nonumber\\
&&+\frac{8}{3}v_3^2\delta_2-\frac{368}{27}v_3\delta_1^2-\frac{368}{27}v_3\delta_2^2
+\frac{92}{27}v_1v_3\delta_1\nonumber\\
&&+\frac{92}{27}v_2v_3\delta_2+8v_1v_3\delta_3
+8v_2v_3\delta_3-64v_3\delta_3^2-64v_3\delta_1\delta_3\nonumber\\
&&-64v_3\delta_2\delta_3,\nonumber\\
&&\beta_{\delta1}=-\delta_1+16\delta_1^2-v_1\delta_1-\frac{1}{3}v_3\delta_3
-\frac{3040}{27}\delta_1^3+\frac{2}{27}v_3^2\delta_3\nonumber\\
&&-\frac{8}{3}v_3\delta_3^2
-\frac{400}{27}v_1\delta_1^2+\frac{23}{81}v_1^2\delta_1
+\frac{5}{243}v_3^2\delta_1-\frac{184}{81}v_3\delta_1\delta_3,\nonumber\\
&&\beta_{\delta2}=-\delta_2+16\delta_2^2-v_1\delta_2-\frac{1}{3}v_3\delta_3
-\frac{3040}{27}\delta_2^3+\frac{2}{27}v_3^2\delta_3\nonumber\\
&&-\frac{8}{3}v_3\delta_3^2
-\frac{400}{27}v_2\delta_2^2+\frac{23}{81}v_2^2\delta_2
+\frac{5}{243}v_3^2\delta_2-\frac{184}{81}v_3\delta_2\delta_3,\nonumber\\
&&\beta_{\delta3}=-\delta_3+8\delta_3^2+\frac{1}{2}v_1\delta_3+\frac{1}{2}v_2\delta_3
+\frac{1}{6}v_3\delta_1+\frac{1}{6}v_3\delta_2\nonumber\\
&&+4\delta_1\delta_3+4\delta_2\delta_3-\frac{64}{3}\delta_3^3+4v_1\delta_3^2
+4v_2\delta_3^2\nonumber\\
&&+\frac{23}{162}v_1^2\delta_3+\frac{23}{162}v_2^2\delta_3
+\frac{368}{27}\delta_1^2\delta_3+\frac{368}{27}\delta_2^2\delta_3\nonumber\\
&&+32\delta_1\delta_3^2+32\delta_2\delta_3^2
+\frac{1}{27}v_3^2\delta_1
+\frac{1}{27}v_3^2\delta_2-\frac{4}{9}v_3\delta_1^2\nonumber\\
&&-\frac{4}{9}v_3\delta_2^2
+\frac{5}{243}v_3^2\delta_3-\frac{40}{81}v_3\delta_3^2
-\frac{92}{27}v_1\delta_1\delta_3-\frac{92}{27}v_2\delta_2\delta_3\nonumber\\
&&-\frac{16}{9}v_3\delta_1\delta_3-\frac{16}{9}v_3\delta_2\delta_3,\nonumber
\end{eqnarray}
\begin{eqnarray}
&&\beta _{z1}=-z_1+v_1z_1+2z_1^3-16\delta_1z_1-4\delta_3z_2+2z_1z_2^2\nonumber\\
&&+\frac{1}{3}v_3z_2-\frac{23}{81}v_1^2z_1-\frac{7}{243}v_3^2z_1-\frac{2}{27}v_3^2z_2\nonumber\\
&&+\frac{29}{27}v_1z_1\delta_1-\frac{736}{27}z_1\delta_1^2-\frac{16}{3}z_1\delta_3^2
-\frac{32}{3}z_2\delta_3^2\nonumber\\
&&+\frac{512}{27}v_3z_1\delta_3+\frac{512}{27}v_3z_2\delta_3,\nonumber\\
&&\beta_{z2}=-z_2+v_2z_2+2z_2^3-16\delta_2z_2-4\delta_3z_1+2z_2z_1^2\nonumber\\
&&+\frac{1}{3}v_3z_1-\frac{23}{81}v_2^2z_2-\frac{7}{243}v_3^2z_2-\frac{2}{27}v_3^2z_1\nonumber\\
&&+\frac{29}{27}v_2z_2\delta_2-\frac{736}{27}z_2\delta_2^2-\frac{16}{3}z_2\delta_3^2
-\frac{32}{3}z_1\delta_3^2\nonumber\\
&&+\frac{512}{27}v_3z_2\delta_3+\frac{512}{27}v_3z_1\delta_3,\nonumber\\
&&\beta _{w1}=-w_1+v_1w_1+2z_1^2w_1-2w_1^3-16\delta_1w_1-4\delta_3w_2\nonumber\\
&&+2w_1z_2^2+\frac{1}{3}v_3w_2-\frac{23}{81}v_1^2w_1
-\frac{7}{243}v_3^2w_1-\frac{2}{27}v_3^2w_2\nonumber\\
&&+\frac{29}{27}v_1w_1\delta_1-\frac{736}{27}w_1\delta_1^2-\frac{16}{3}w_1\delta_3^2
-\frac{32}{3}w_2\delta_3^2+\nonumber\\
&&+\frac{512}{27}v_3w_1\delta_3+\frac{512}{27}v_3w_2\delta_3,\nonumber\\
&&\beta _{w2}=-w_2+v_2w_2+2z_2^2w_2-2w_2^3-16\delta_2w_2-4\delta_3w_1\nonumber\\
&&+2w_2z_1^2+\frac{1}{3}v_3w_1-\frac{23}{81}v_2^2w_2
-\frac{7}{243}v_3^2w_2-\frac{2}{27}v_3^2w_1\nonumber\\
&&+\frac{29}{27}v_2w_2\delta_2
-\frac{736}{27}w_2\delta_2^2-\frac{16}{3}w_2\delta_3^2-\frac{32}{3}w_1\delta_3^2-\nonumber\\
&&+\frac{512}{27}v_3w_2\delta_3+\frac{512}{27}v_3w_2\delta_3.\nonumber
\end{eqnarray}
It is well known that the perturbative series expansions are
asymptotic, and the vertices of the interactions of the order
parameter fluctuations in the fluctuation region are sufficiently
large for Eqs. (6) to be directly applied. Therefore, to extract
the necessary physical information from the expressions derived
above, the Pade-Borel method generalized to the multiparameter
case was used. The corresponding direct and inverse Borel
transformations have the form
\begin{eqnarray}
&& f(v_1,v_2,v_3,\delta_1,\delta_2,\delta_3,z_1,z_2,w_1,w_2)\\
&&=\sum\limits_{i_1,...,i_{10}}
c_{i_1...i_10}v_1^{i_1}v_2^{i_2}v_3^{i_3}\delta_1^{i_4}\delta_2^{i_5}\delta_3^{i_6}
z_1^{i_7}z_2^{i_8}w_1^{i_9}w_2^{i_{10}}\nonumber\\
&&=\int\limits_{0}^{\infty}e^{-t}F(v_1t,v_2t,v_3t,\delta_1t,\delta_2t,\delta_3t,z_1t,z_2t,w_1t,w_2t)dt,\nonumber\\
&& F(v_1,v_2,v_3,\delta_1,\delta_2,\delta_3,z_1,z_2,w_1,w_2)\nonumber\\
&&=\sum\limits_{i_1,...,i_{10}}\frac{\displaystyle c_{i_1,...,i_7}}
{\displaystyle(i_1+...+i_{10})!}v_1^{i_1}v_2^{i_2}v_3^{i_3}\delta_1^{i_4}\delta_2^{i_5}\delta_3^{i_6}
z_1^{i_7}z_2^{i_8}w_1^{i_9}w_2^{i_{10}}.\nonumber
\end{eqnarray}
For an analytic continuation of the Borel transform of a function,
a series in an auxiliary variable $\theta$ is introduced:
\begin{eqnarray}
&&\tilde{F}(v_1,v_2,v_3,\delta_1,\delta_2,\delta_3,z_1,z_2,w_1,w_2,\theta)
=\sum\limits_{k=0}^{\infty}\theta^k\times\nonumber\\
&&\times\sum\limits_{i_1,...,i_{10}}
\frac{\displaystyle c_{i_1...i_10}}{\displaystyle k!}
v_1^{i_1}v_2^{i_2}v_3^{i_3}\delta_1^{i_4}\delta_2^{i_5}\delta_3^{i_6}
z_1^{i_7}z_2^{i_8}w_1^{i_9}w_2^{i_{10}}\delta_{i_1+...+i_{10},k},\nonumber
\end{eqnarray}
and the [L/M] Pade approximation is applied to this series at the
point $\theta=1$. This approach was proposed and tested in [9] for
describing the critical behavior of systems characterized by
several vertices corresponding to the interaction of the order
parameter fluctuations. The property [9] that the system retains
its symmetry under the Pade approximants in the variable $\theta$ is
essential in the description of multivertex models.

In the two-loop approximation, the â functions were calculated
using the [2/1] approximant. The character of the critical
behavior is determined by the existence of a stable fixed point
satisfying the set of equations
\begin{eqnarray}
&&\beta_{i}(v_1^*,v_2^*,v_3^*,\delta_1^*,\delta_2^*,\delta_3^*,z_1^*,z_2^*,w_1^*,w_2^*)=0\nonumber\\
&&(i=1,...,10).
\end{eqnarray}
The requirement that the fixed point be stable is reduced to the
condition that the eigenvalues $b_i$ of the matrix
\begin{eqnarray}  \displaystyle
&&B_{i,j}=\frac{\partial\beta_i(v_1^*,v_2^*,v_3^*,\delta_1^*,\delta_2^*,\delta_3^*,z_1^*,z_2^*,w_1^*,w_2^*)}{\partial{v_j}}\nonumber\\
&&(v_i,v_j \equiv v_1^*,v_2^*,v_3^*,\delta_1^*,\delta_2^*,\delta_3^*,z_1^*,z_2^*,w_1^*,w_2^*)
\end{eqnarray}
lie on the right-hand complex half-plane.

The resulting set of summed $\beta$ functions contains a wide variety
of fixed points lying in the physical region of the vertex values
with $v_i\geq 0$.

A complete analysis of the fixed points, each of which corresponds
to the critical behavior of a single order parameter, was
presented in our recent publication [5]. Now, we consider the
combined critical behavior of two order parameters.

The analysis of the values and stability of the fixed points
offers a number of conclusions. The tetracritical fixed point of a
disordered incompressible system ($v_1=v_2=1.58892$,
$v_3=0$, $\delta_1=\delta_2=0.03448$, $\delta_3=0$, $z_1=0$,
$z_2=0$, $w_1=0$, $w_2=0$) is unstable
under the effect of uniform deformations ($b_1=b_2=0.461$, $b_3=0.036$,
$b_4=b_5=0.461$, $b_6=0.036$, $b_7=b_8=b_9=b_{10}=-0.236$).
The striction effects lead to the stabilization of the
tetracritical fixed point of a compressible disordered system
($v_1=v_2=1.58892$, $v_3=0$,$\delta_1=\delta_2=0.03448$, $\delta_3=0$,
$z_1=0.04599$, $z_2=0.568836$, $w_1=0.017759$, $w_2=00.551849$,
$b_1=b_2=0.461$, $b_3=0.036$, $b_4=b_5=0.461$, $b_6=0.036$,
$b_7=1.189$, $b_8=0.003$, $b_9=5.391$, $b_7=0.999$).

The question about the stability of other multicritical points
cannot be resolved in terms of the described model, because the
calculations lead to a degenerate set of equations. The degeneracy
can be removed by considering the Hamiltonian with allowance for
the terms of higher orders in both the strain tensor components
and the fluctuating order parameters.

Thus, the striction-induced interaction of the fluctuating order
parameters with elastic deformations, as well as the introduction
of frozen point impurities into the system, leads to a change from
the bicritical behavior to the tetracritical one. Elastic
deformations lead to a change in the regime of tetracritical
behavior of a disordered system because of the interaction between
the order parameters through the acoustic phonon exchange.

The work is supported by Russian Foundation for Basic Research N
04-02-16002.

\def\baselinestretch{1.0}

\end{document}